\documentclass[
    ,final            
    ,natbib
  ]
  {aipproc}

\layoutstyle{8x11double}

\newcommand{\xte}{{\it RXTE}}

\newcommand{\epcs}{{\rm erg\,cm^{-2}\,s^{-1}}}
\newcommand{\epc}{{\rm erg\,cm^{-2}}}

\newcommand{\gtrsim}{\mathrel{\hbox{\rlap{\hbox{\lower4pt\hbox{$\sim$}}}\hbox{$>$}}}}
\newcommand{\lesssim}{\mathrel{\hbox{\rlap{\hbox{\lower4pt\hbox{$\sim$}}}\hbox{$<$}}}}
\newcommand{\apj}{ApJ}
\newcommand{\apjl}{ApJL}
\newcommand{\apjs}{ApJS}
\newcommand{\aap}{A\&A}
\newcommand{\mnras}{MNRAS}
\newcommand{\nat}{Nature}
\newcommand{\physrep}{Phys.~Rep.}
\newcommand{\prd}{Phys.~Rev.~D}
\newcommand{\ssr}{Space Sci.~Rev.}
\newcommand{\iaucirc}{IAU Circ.}
\newcommand{\procspie}{Proc. SPIE}

\newcommand{\nmspsmo}{seven}	
\newcommand{\numns}{22}

\newcommand{\srca}{XTE~J1814$-$338}
\newcommand{\srcb}{SAX~J1808.4$-$3658}

\newcommand{\srcd}{XTE~J1751$-$305}

\newcommand{\srcf}{IGR~J00291+5934}
\newcommand{\srcg}{HETE~J1900.1$-$2455}
\newcommand{\srch}{Swift~J1756.9$-$2508}

\begin{document}

\title{Accreting neutron star spins and the equation of state}


\author{Duncan Galloway}{
  address={School of Physics \& School of Mathematical Sciences, Monash
University, VIC 3800, Australia}
}

\begin{abstract}
X-ray timing of neutron stars in low-mass X-ray
binaries (LMXBs) with the {\it Rossi X-ray Timing Explorer}\/ has since 1996
revealed several distinct high-frequency phenomena. Among these are
oscillations during thermonuclear (type-I) bursts, which (in addition to
persistent X-ray pulsations) are thought to trace the neutron star spin.
The recent discoveries of 294~Hz burst oscillations in IGR~J17191$-$2821,
and 182~Hz pulsations in Swift~J1756.9$-$2508, brings the total number of
measured LMXB spin rates to 22.
An open question is why the majority of the $\approx100$ known neutron
stars in LMXBs show neither pulsations nor burst oscillations. 

Recent
observations suggest
that persistent pulsations may be more common than previously thought,
although detectable intermittently, and in some cases at very low duty cycles.
For example, the 377.3~Hz pulsations in 
HETE~J1900.1$-$2455 were only present in the first
few months of it's outburst, and have been absent since (although X-ray
activity continues).
Intermittent (persistent) pulsations 
have since been detected in a further two sources. In two of these three
systems the pulsations appear to be related to the thermonuclear burst
activity, but in the third (Aql~X-1) they are not.
This phenomenon offers new opportunities for spin measurements in known
systems.

Such measurements 
can constrain the poorly-known neutron star equation of state,
and neutron stars in LMXBs offer observational advantages over
rotation-powered pulsars which make the detection of more rapidly-spinning
examples more likely. 
Even so,
spin rates of at least 
50\% faster than the present maximum appear necessary to give constraints
stringent enough to discriminate between the various models.
Although the future prospects for such rapidly-spinning objects do not appear
optimistic, several additional observational approaches are possible for
LMXBs.
The recent study of EXO~0748$-$676 is an example. 
\end{abstract}

\maketitle

\section{Introduction}

The equation of state (EOS) of cold matter at super-nuclear densities
remains one of the foremost outstanding problems for fundamental physics
(e.g. \citep{lp07}).
The major uncertainty in 
high-density quantum chromo-dynamics 
theory (which has otherwise been so successful in explaining the
properties and behaviour of subatomic particles) is in the regime where
the density is at or above that reached in the atomic nucleus. Cold matter
beyond nuclear density may be composed primarily of neutrons, as is
normally thought, or it could be dominated by exotic components such as
hyperons, pion or kaon condensates, or quark matter
(e.g. \citep{witten84}).
Such states of matter are {\em purely theoretical} at the present time,
and their detection --- whether it be 
via accelerator experiments, or in
the astrophysical ``laboratories'' available to astronomers ---
would represent a significant step forward for modern physics.

Particle accelerators probe the conditions in matter at extreme
temperatures and
densities (up to a factor of ten higher than nuclear). 
Matter within neutron stars is also expected to reach super-nuclear
densities, but at comparatively ``cool'' temperatures (no more than
$10^9$~K!). Neutron stars thus play an important complementary role for
studies of condensed matter, and measurements which may constrain the EOS
are a high
priority for observers.

Since the EOS affects the bulk properties (mass and radius) of neutron stars,
simultaneous measurement of these parameters with moderate precision in an
individual object would in some cases be sufficient to identify the EOS.
However, such measurements have proved surprisingly elusive.
The masses of neutron stars in binary (rotation-powered) pulsars can be
measured in some cases to a fraction of a percent (e.g. \citep{tc99})
although simultaneous radius measurements are generally not available.
While the maximum neutron star mass also provides a constraint on the EOS,
most of the measured masses cluster around $1.4\,M_\odot$, which is not
useful in distinguishing between different models.

Measurement of the spin rate in rapidly-rotating neutron stars provides
a relatively model-independent way to constrain the EOS.
The maximum spin rate of a neutron star (above
which it will break up due to centrifugal forces) can be expressed in
terms of the neutron star mass $M$ and radius $R$, roughly independent
of the 
EOS
\cite[]{lp07}: 

\begin{equation}
\nu_{\rm max} = 1045(M/M_\odot)^{1/2}(10\,{\rm km}/R)^{3/2}~{\rm Hz}
\end{equation}
where $M$ and $R$ are the neutron star mass and radius.
Constraining the possible candidates for the neutron star EOS thus
requires detection of ever-more rapidly spinning neutron stars.
The fastest-spinning neutron star presently known is the radio pulsar
PSR~J1748$-$2446ad, at 716~Hz \cite[]{hessels06}.
Although
it's mass 
is unknown, 
assuming a value consistent with the measurements from other radio
pulsars leads to an upper radius limit 
of 14.4~km. Without a mass
measurement, this limit does not yet allow us to reject any individual
EOS.

A compelling observational goal, then, is to detect ever-more rapidly
spinning neutron stars.  Despite much effort, progress in this regard has
been slow; the detection of PSR~J1748$-$2446ad represented the first
increase in the known maximum neutron-star spin rate in 23 years. In the
radio band, selection effects make more rapidly-spinning rotation powered
pulsars significantly harder to detect.
Although these selection effects do not affect spin measurements via X-ray
observations of accretion-powered neutron stars, faster-spinning systems
have not yet been convincingly detected, suggesting perhaps that some
physical mechanism prevents further spin-up.

Regardless, measurement of neutron-star spins in LMXBs remains a high
observational priority. In this paper I will discuss the phenomenology of
the various types of high-frequency timing phenomena detected to date, and
assess the prospects for future detections which may provide the first
strong constraints on the neutron-star equation of state.

\section{Measurement of NS spins in LMXBs}

Evidence of rapid spins in neutron-star LMXBs has been obtained
exclusively via observations with the Proportional Counter Array
(PCA; \citep{xte96}) aboard the {\it Rossi X-ray Timing Explorer}\/
({\it RXTE}). The PCA is the only instrument to date with the sensitivity
(effective area $\approx6500$~cm$^2$) and time resolution ($\approx1\
\mu$s) necessary to detect rapid variability from these systems.
With the exception of a few high-field neutron stars in LMXB systems
(including Her~X-1 and GX~1+4), measured spin frequencies fall in the
range 45--620~Hz (Table \ref{tab:a}), with all but one $>180$~Hz.
These rapid spins confirm the LMXBs as the evolutionary precursor to
the ``recycled'' millisecond radio pulsars \cite[]{alpar82,rs82}.  

The LMXBs for which spins have been measured represent only about 20\% of
the known population of more than 100 (e.g. \citep{lmxb07}).  It remains
an open question as to why it is so difficult to measure the spin in the
majority of neutron stars in LMXBs. The two conventional explanations are
that either the non-pulsing neutron stars in LMXBs have magnetic fields
that are too weak to channel accretion onto polar hotspots (perhaps
due to suppression by the accreted material; e.g. \citep{czb01}) or that the
pulsations are scattered by a surrounding region of high optical depth
(e.g. \citep{tcw02}). A comparison of the spectral properties of the
pulsing and non-pulsing LMXBs does not support the latter explanation
(\citep{krauss04,gogus07}, although see also \citep{tks07}). Furthermore,
while the sources which exhibit pulsations tend to have low time-averaged
X-ray fluxes (and hence accretion rates), this condition is not sufficient
for pulsations to be detectable.
The contrast with the rotation-powered pulsars is even more marked when
one considers that even the LMXBs which {\it do}\/ exhibit pulsations, do
not exhibit pulsations at all times. Pulsations may only be detected from
the accretion-powered millisecond pulsars (AMSPs) when in outburst;
similarly, burst oscillations are only detected for a few seconds at the
peak of some thermonuclear bursts. This property presents an observational
challenge to the measurement of rapid neutron star spins which is quite
distinct from the difficulties encountered in searches for
rapidly-spinning rotation-powered pulsars.

In further contrast to the rotation-powered pulsars, the spin rate for
neutron stars in LMXBs may be measured in two distinct
ways\footnote{Here I exclude measurement of the separation frequency of
pairs of high-frequency quasi-periodic oscillations, which has long
been thought to be approximately equal to or half the spin frequency
(although see \citep{mb07}). }:
burst oscillations and persistent pulsations.
In addition, intermittent (persistent) pulsations have been detected
recently in three systems.
Below I describe each of these phenomena
in more detail.

\subsection{Burst oscillations}
\label{bo}

The presence of X-ray bursts are practically a defining characteristic of
LMXBs (e.g. \citep{lew93,sb03}). Thermonuclear (type-I) bursts are caused
by unstable ignition of accumulated H/He on the surface of accreting
neutron stars; the X-ray intensity increases by an order of magnitude
within at most a few seconds, before decreasing back to the persistent
level within 10--100~s. Although there are several observational aspects
which continue to defy explanation (e.g. \citep{bcatalog}), the physics of
the nuclear ignition and burning are reasonably well understood, and in
some cases are fully consistent with observations \cite[]{gal03d}.

Rapid (363~Hz) oscillations were first discovered in bursts from the
well-known persistent X-ray source 4U~1728$-$34 \cite[]{stroh96}. A
power-density spectrum covering the maximum of the burst exhibited
multiple closely-spaced peaks, that were later resolved into a continuous
upwards frequency drift over the span of the oscillation. Frequency
drifts of a few Hz, as well as $\approx10$\% amplitudes and 
sinusoidal pulse profiles, subsequently proved to be typical of such
oscillations.  The high coherence of these oscillations long recommended
them as tracers of the neutron star spin, and this conjecture was all but
confirmed with the detection of burst oscillations at the persistent
pulsation frequency in two accretion-powered pulsars
\cite[]{chak03a,stroh03a}.

Burst oscillations have been discovered to date in 14 sources
(e.g. \citep{sb03}; Table \ref{tab:a}), at frequencies in the range
45--620~Hz. The oscillations with the lowest frequency were detected by
summing power-density spectra of 38 bursts detected from EXO~0748$-$676
\cite[]{villarreal04}. In this source, the oscillations are uniquely not
detectable in individual bursts. The highest frequency oscillations to
date are from 4U~1608$-$52, at 620~Hz (Hartman et al., 2008, in
preparation). 
The most recent discovery has been in the rapid transient
IGR~J17191$-$2821. A thermonuclear burst was detected from this system by
\xte/PCA on 2007 May 4, in which high-frequency oscillations were present,
increasing in frequency from 292 to 294~Hz \cite[]{mark07a}. As with the
other burst oscillation sources, the highest freqency detected is assumed
to be the neutron star spin frequency. 

Both the average and maximum spin frequencies of the
burst oscillation sources are higher than those of the sources with
persistent pulsations, so that this phenomenon perhaps offers the best
opportunity for increasing the maximum spin rate for rapidly-rotating
neutron stars.
Evidence for a burst oscillation at $>1000$~Hz has already been
reported, although the low significance of the signal means that it must
be considered a candidate, at best.
A peak at 1122~Hz was detected in the power-density spectrum of a 4-s
interval late in the tail of a thermonuclear burst from the LMXB transient
XTE~J1739$-$285 \cite[]{kaaret07a}.
However, 
no comparable power excess was detected at this frequency 
in other (non-overlapping) intervals during the burst, nor in any of the
other six bursts observed by {\it RXTE}.  Furthermore, while the
significance of the signal was estimated at $3.97\sigma$ based on
Monte-Carlo simulations, a standard
calculation taking into account the total number of trials (for 
overlapping 4-s time windows up to the Nyquist frequency) suggests the
significance is at most $3.5\sigma$.
At this relatively low significance, without corroborating detections in
other bursts from this source (or at least in other independent,
non-overlapping time intervals) this 
detection cannot yet be interpreted as a spin measurement.

\subsection{Persistent X-ray pulsations}
\label{pers}

The accretion-powered millisecond pulsars (AMSPs) have
emerged as a distinct class of LMXBs, beginning with the
discovery of pulsations in SAX~J1808.4$-$3658 in 1998
\cite[]{wij98b,chak98d}. Since then, \nmspsmo\ more AMSPs (including
\srcg, which is more accurately classified as an intermittent pulsar,
below) were discovered during
transient outbursts typically lasting a few weeks
(see \citep{wij04a} for a review).
Extensive observations with \xte\/ and other instruments
have revealed a number of properties largely
characteristic of the class. The outbursts tend to be of short duration,
typically a few weeks
(but as long as 50~d in XTE~J1814$-$338).
Pulsations are persistently detected at fractional amplitudes of typically
$\sim5$\% rms.  Where thermonuclear bursts are present, oscillations at
the pulsation frequency and roughly the same fractional amplitude are
present throughout (e.g. \citep{chak03a,stroh03a}).

The most recently-discovered source, Swift~J1756.9$-$2508, is an exemplar
of the class. This system was discovered when it began a transient outburst
and was detected by the Burst Alert Telescope (BAT) aboard the {\it
Swift}\/ satellite on 2007 June 7 \cite[]{krimm07}. A subsequent \xte\/
observation of the field showed a significant excess in the power-density
spectrum at 182~Hz, confirming the source as an accretion-powered
millisecond pulsar. Pulse timing of subsequent observations precisely
measured Doppler delays from a 54.7~min binary orbit. The Roche
lobe in such an ``ultracompact'' binary cannot accommodate a main-sequence
donor, and the likely companion is He-dominated, with a mass in the range
(6.7--$22)\times10^{-3}\ M_\odot$. Approximately 13 days later the X-ray
flux had dropped to several orders of magnitude below the outburst maximum,
and the system had all but returned to quiescence. Searches for X-ray
emission from the source over the preceding 2.5~yr for which BAT data was
available, as well as the 11.4~yr interval spanned by \xte/PCA and ASM
measurements, were unsuccessful (although the sensitivity to faint
outbursts is reduced due to the nearby bright source GX~5-1). This
suggests that the outburst recurrence time for Swift~J1756.9$-$2508 is
$\gtrsim10$~yr, similar to the other ultracompact AMSPs.

The characteristic short-duration outbursts coupled with recurrence times
of years result in low time-averaged accretion rates for the AMSPs, of
order $10^{-11}\ M_\odot\,{\rm yr^{-1}}$ \cite[]{gal06b}. Five of the
eight known systems have been detected only once in outburst, so that the
actual recurrence time is unknown.
The three systems which have
exhibited multiple outbursts, exhibit two distinct
recurrence patterns. First, in \srcd, a strong (maximum 50~mCrab) outburst
(which led to the source discovery) in 2002 was followed by two much
shorter and weaker ($\approx20$~mCrab) outbursts, 3 and 2 years later 
\cite[]{greb05a,lin07a}.
The estimated fluence from the latest mini-outburst, in
2007, allows a rough measure of the time-averaged flux of
$1.6\times10^{-12}\ \epcs$, at least an order of magnitude smaller than
that of the other AMSPs (excluding possibly \srch).

The second characteristic pattern of outbursts is typified by the
behaviour of
\srcb.
To date, five outbursts with comparable durations, peak intensities, and
fluences have been observed, that were separated by $2.2\pm0.6$~yr on
average \cite[]{gal06b}
The
similarity of the outburst profiles extends to large-amplitude variations
in X-ray flux for $\approx15$~d prior to the transition to quiescence
(see e.g. \citep{gal06c}), as well as the pattern of X-ray pulse
variation \cite[]{hartman07}.
For \srcf, the system which is most similar in its system properties to
\srcb, a retroactive search of the ASM intensity history revealed
evidence for two previous outbursts, 3 and 6~yr earlier. 
The variability in the outburst intervals for
\srcf\ was substantially less, although 
a fluence measurement was possible only for the latest outburst, so that
the degree of variation of the long-term accretion rate from interval to
interval is unknown.

\begin{figure}
 \includegraphics[width=0.45\textwidth]{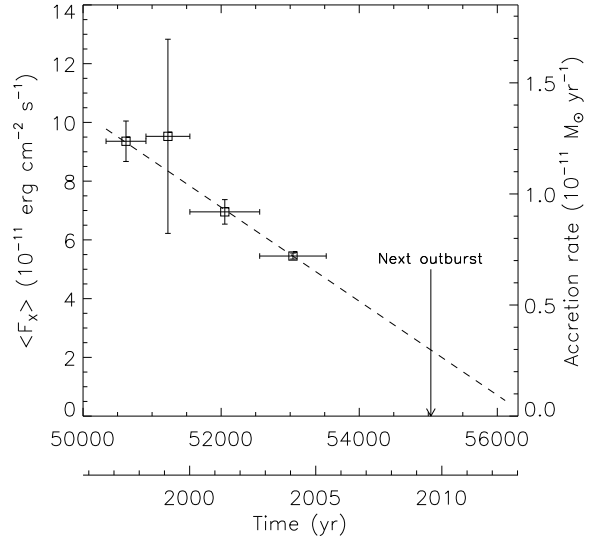}
 \caption{Evolution of the long-term time-averaged flux in \srcb.
Each measurement is based on the interval between outbursts and the
fluence measured for the outburst which followed, 
as calculated by \cite[]{gal06b} but using an improved measure of the
outburst fluence for the 2005 June outburst (see text).
The right-hand $y$-axis indicates the corresponding mass
accretion rate for a $1.4\ M_\odot$ neutron star with $R=10$~km at
3.6~kpc. The dashed line is a linear line of best fit, projected through
the implied time for the next outburst.
  \label{projection} }
\end{figure}

In an earlier study the outburst fluences for \srcb\ were found to be
roughly similar, although the fluence measurement for the 2005 June
outburst was based on ASM data only, since no public PCA data were
available \cite{gal06b}. The now-public PCA observations, which offer
excellent coverage of the outburst (even including the rise) allow a much
more precise measure of the fluence, of $(4.54\pm0.08)\times10^{-3}\ \epc$.
With this more precise measure, I find that the outburst fluences
deviate from a constant value at the $5\sigma$ level. These variations in
the outburst fluence, coupled with the significant variations
in the outburst interval, have the consequence that
the mean accretion rate in \srcb\ has decreased by about 40\%
between 1996-1998 and 2002-2005. 
The mean flux (and accretion rate) plotted for each outburst clearly show
a steadily decreasing trend (Figure \ref{projection}).

\begin{figure}
 \includegraphics[width=0.45\textwidth]{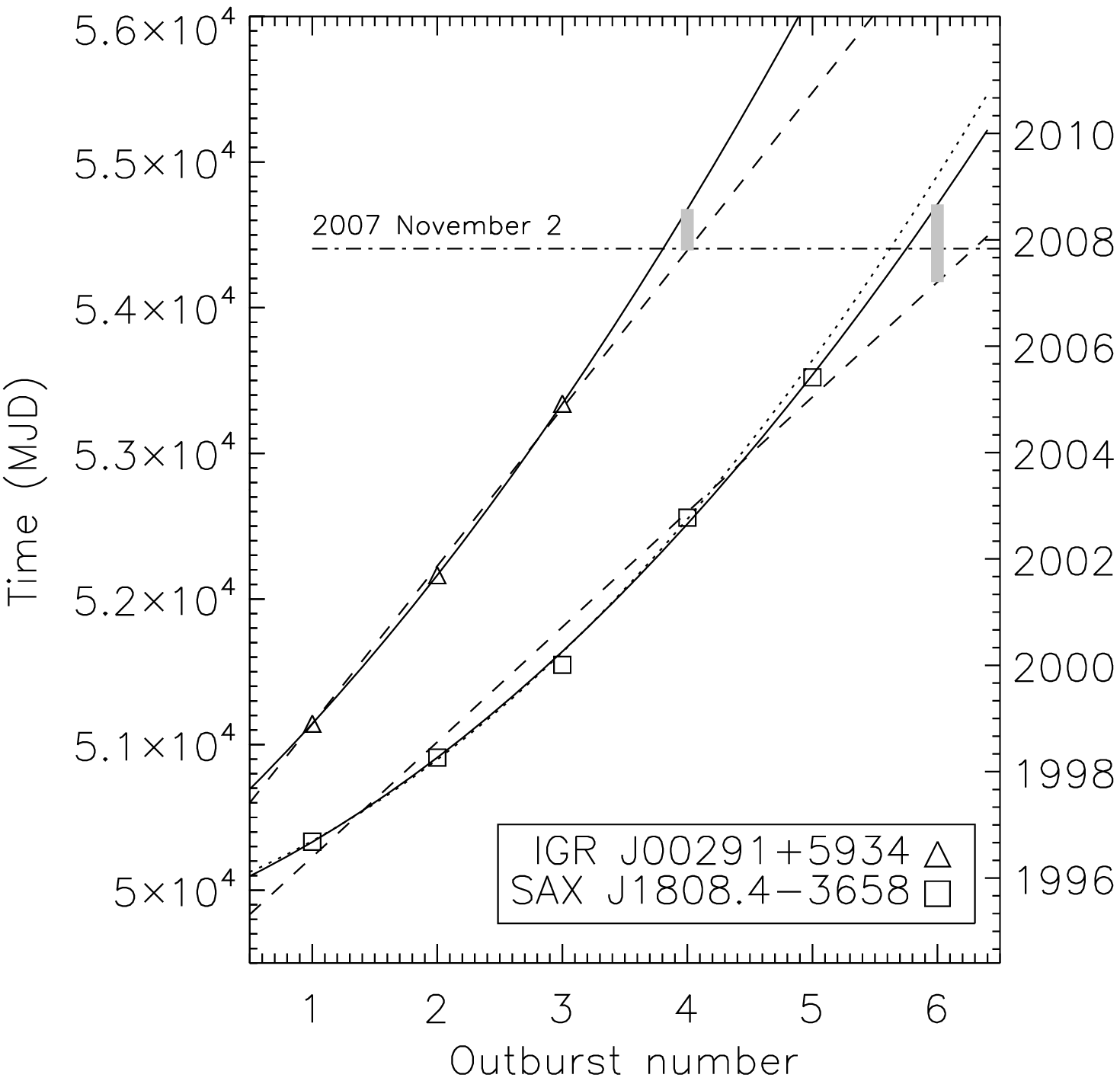}
 \caption{Outburst times and phenomenological model fits for the recurrent
transients \srcb\ and \srcf. The open symbols show the actual start of
each outburst detected by {\it BeppoSAX}\/, the \xte/ASM or PCA.
Linear ({\it dashed lines}) and quadratic ({\it
continuous lines}) fits to the recurrence times for each source are shown.
For \srcb, I also show the quadratic fit derived prior to the 2005 June
outburst ({\it dotted line}), which was accurate in predicting the time of
that occurrence to within 12\% (of the outburst interval). The predicted
time ranges for the next outburst in each source, calculated as the span
of the linear and quadratic models, is shown as the filled grey box.
 \label{plots} }
\end{figure}

The reliability of predictions for subsequent outbursts in these repeating
transients is an important factor for observers, not only in the X-ray
band.
In 2004 December, following an analysis of the recurrence times of the
outbursts of \srcb\ and \srcf\ observed until then, I compared predictions of
linear and quadratic fits of the outburst recurrence time. The
quadratic fit to the outburst times for \srcb\ resulted in much smaller
residuals, and predicted the next outburst in 2005 September--October. The
outburst actually occured three months earlier, in 2005 June, an error of
just 12\% of the recurrence time. The early occurrence of this outburst
compared to the prediction may have been related to the fact that the
outburst fluence was the smallest yet measured for \srcb\ 
\cite[]{gal06b}.
Encouraged by the success of the phenomenological model fits in predicting
the 2005 June outburst, I make further predictions for the next outbursts
in both \srcb\ and \srcf. For \srcf, the projections of the linear and
quadratic fits do not diverge substantially through the time of the next
outburst (Fig. \ref{plots}). The time range spanned by the two models are
MJD~54390--54680, i.e. between 2007 October and 2008 July.
For \srcb, the divergence between the linear and quadratic models is more
significant, and in fact the linear model predicts the time for the next
outburst as early as 2007 March\footnote{Since the X-ray observational
coverage of the Galactic bulge region (which includes \srcb) is better
than anywhere else in the sky, we can confidently rule out the sixth
outburst having already occurred}. Thus, I predict the next outburst to
occur sometime between 2007 September and 2008 July.

An alternative prediction for the next outburst in \srcb\ may be made
based on the trend of the long-term time-averaged flux.
Extrapolating a linear fit to the measurements suggests that sufficient material
will have been accreted to the disk to power an outburst of fluence equal
to that of 2005 June by MJD~55040 (2009 July; Fig. \ref{projection}). Thus, 
an outburst of fluence less than or equal to that in 2005 June will likely
occur no later than 2009 July.
Interestingly, extrapolating the linear trend further
suggests that accretion will cease
altogether by around 2013, although this seems implausible!

The wide-field instruments onboard {\it INTEGRAL}, {\it Swift}\/ and (to a
lesser extent) \xte\/ and {\it HETE-II}\/ have resulted in a discovery
rate for these systems of about 1.4~yr$^{-1}$ since 2002. That this level
of coverage has been sustained now for 5~yr suggests that the sample of
short ($\approx2$--3~yr) recurrence time transient AMSPs is practically
complete. If this is indeed the case, the future discoveries are likely to
be systems with substantially longer recurrence times (such that they have
not yet been in outburst since 2002, or even earlier). There are already
indications that the discovery rate for AMSPs is decreasing 
with time. Swift~J1756.9$-$2508, discovered in 2007 June, was the first
transient AMSP discovered in 2.5~yr. The new sources discovered in the
future, as well as observations of repeat outbursts of the known systems,
will be critical to constrain the presently highly uncertain distribution
of recurrence times (and hence accretion rates) of these systems.

\subsection{Intermittent pulsations}
\label{int}

Perhaps no less puzzling than the question of why persistent pulsations
were only detected in the handful of AMSPs prior to 2006, was the fact
that the division between the two classes of systems --- the AMSPs and the
non-pulsing LMXBs --- was so sharp. Despite deep searches by a number of
observers, persistent pulsations had not been detected in any other LMXBs,
even when a measured burst oscillation frequency could be used to guide
the search. Conversely, the pulsations in the first six AMSPs discovered
were always present when the sources were detectable with \xte/PCA. This
division has since been weakened by the detection of intermittent
(persistent) pulsations, first in the long-duration transient
HETE~J1900.1$-$2455, and subsequently in two additional sources.

HETE~J1900.1$-$2455 was discovered when a bright thermonuclear (type-I)
burst was detected with the {\it HETE-II}\/ satellite on 2005 June 14, and
a subsequent \xte/PCA observation revealed the presence of 377~Hz
pulsations \cite[]{kaaret05b}. Pulse timing of the observations which
followed revealed Doppler shifts from an 83.3~min binary orbit, with a
companion likely having mass in the range (16--$70)\times10^{-3}\ M_\odot$.
This system soon revealed several properties distinct from the population
of AMSPs known until that time. First, the system remained active long
after the usual outburst duration for the AMSPs, and in fact is still
active (as of 2007 November) at approximately the same mean X-ray flux
since 2005 June. Second, the pulse amplitude was unusually low (at most
3\%~rms), and decreased on a 10-d timescale following several
thermonuclear bursts observed early in the outburst \cite[]{gal07a}.
Third, and perhaps most interestingly, was that the pulsations became
undetectable on several occasions in the first few months of the outburst,
and since 2005 August 20 have not been detected at all. Weekly \xte\/ 
observations continue with the goal of detecting any change in the system
flux or the return of pulsations.

The behaviour of pulsations in this system is enigmatic, having a
complex relationship with the presence of thermonuclear bursts. On three
occasions the pulsations appeared strongly close to the times of
thermonuclear bursts, and then decreasing gradually in amplitude until the
next. While this suggests that the bursts themselves triggered the
appearance of the pulsations, and in one case a burst preceded the first
detection in that observation, in another case the detection of pulsations
instead preceded a burst by 2.4~hr. Furthermore, while the source
continued bursting after 2005 August 20, the subsequent bursts were not
accompanied by pulsations at any level.

The phenomenology has become even more complex with the detection of
persistent pulsations in two more systems. Timing analysis of the entire
1.3~Ms of \xte/PCA data accumulated on the well-known transient LMXB
Aql~X-1 resulted in a single detection of persistent pulsations on 1998
March 10, lasting approximately 150~s \cite[]{casella07}. The pulsation,
at a frequency of 550.27~Hz, was $\approx0.53$~Hz higher than the
asymptotic frequency of burst oscillations observed from the source. 
No bursts were observed within several days of the observation that
exhibited pulsations, and no spectral variation was detected while the
pulsations were present.
The estimated duty cycle for the pulsation was just $3\times10^{-4}$.
In the globular cluster LMXB SAX~J1748.9$-$2021, pulsations at 442~Hz were
detected on several occasions in 2001 and 2005
\cite[]{gavriil07,altamirano07}\footnote{SAX~J1748.9$-$2021 had previously been
reported as a 410~Hz burst oscillation source \cite[]{kaaret03}, although
that signal was detected only briefly in a single burst, and at low
significance. While a source with pulsations and burst oscillations at
different frequencies would be truly remarkable, the burst oscillation
detection was likely not real.}. The oscillations were present during an
interval in which several bursts were detected, and 
exhibited Doppler variations in frequency consistent with a binary orbit
with period 8.76~hr.

It may appear an artifical distinction to separate the intermittent
pulsars from the seven other ``classical'' AMSPs, but there are several
other distinguishing characteristics. Most notably, the properties of the
pulsations in two of the three intermittent systems appear to be related
to the occurrence of thermonuclear bursts.  Bursts have also been observed
from \srcb\ and \srca, although with no apparent effect on the persistent
pulsations.  While it seems implausible that a separate pulse emission
mechanism is involved, the mechanism behind the appearance and
disappearance of pulsations in these systems is presently a mystery.
More detailed studies of the pulse and spectral properties in the known
sources, as well as observations of additional examples, may provide the
solution.

\section{Discussion}

\begin{figure}
 \includegraphics[width=0.45\textwidth]{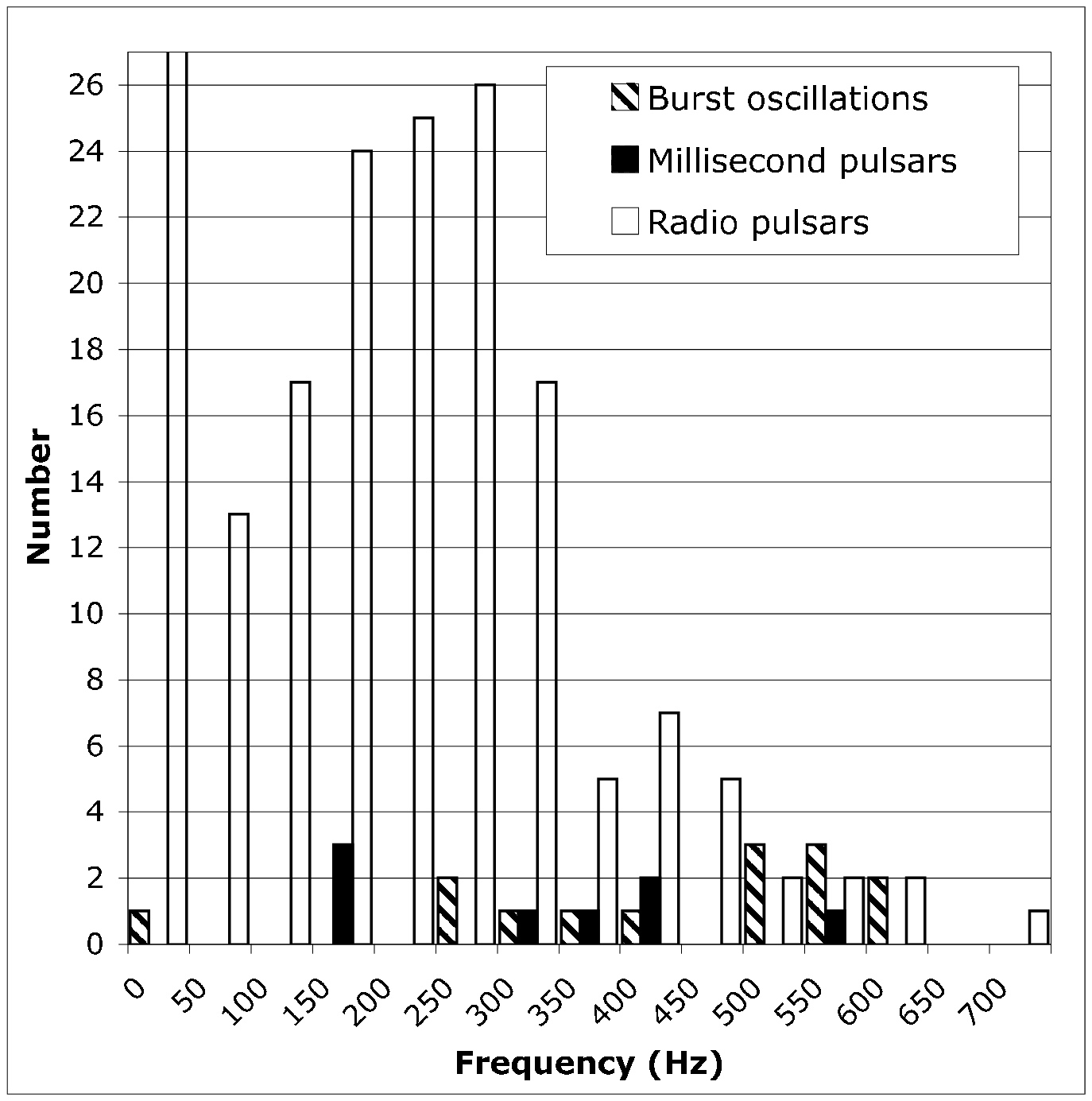}
 \caption{
The neutron star spin frequency distribution,
plotted
separately for different types of systems: radio (rotation-powered)
pulsars (from the ATNF Pulsar Catalogue, as of June 2007),
accretion-powered millisecond pulsars, and burst oscillation sources.
The overwhelming majority of rotation-powered pulsars spin slowly;
the bar at 0--50~Hz is cut off by the $y$-axis range and includes 
1480
sources. The distribution for the accreting sources is much flatter, and
is not affected by any known selection effects which make detection of
more rapidly-spinning systems less likely.
  \label{spins} }
\end{figure}

Having presented in some detail the phenomenology of measuring accreting
neutron-star spins, I turn to the prospects for the future potential for
stringent constraints on the neutron-star equation of state.
The combined spin frequency distribution for the \numns\ burst oscillation
sources and millisecond pulsars is approximately flat between
45--620~Hz (Fig. \ref{spins}).
The spin distribution for rotation-powered pulsars, in contrast, is subject
to significant selection effects which mask the true distribution.
Radio pulsars are subject to pulse smearing due to dispersion in the
interstellar medium, which makes previously unknown 
examples harder to
detect. This problem becomes worse for very fast pulsars, and the 
increasing effect of eclipses by the outflowing pulsar wind
further decreases the
sensitivity for detection. X-ray pulsations, on the other hand, are not
subject to either effect, and
thanks to {\it RXTE}'s
timing capability well above 1~kHz, the sensitivity to X-ray
pulsars rotating at
frequencies well above the current maximum is effectively flat. Thus,
accretion-powered pulsars may offer the best chance to detect
maximally-rotating neutron stars, and thus provide future constraints on the
neutron star EOS.

In the absence of any known selection effects, the present lack of
accretion-powered neutron stars spinning faster than 620~Hz strongly
suggests that such systems 
are rare, if they exist at all. One possible explanation for this lack of
faster-spinning objects is the increasing role of gravitational radiation
which may prevent further spin-up (e.g. \citep{bil98c}). Regardless of the
mechanism which apparently prevents further spin-up,
this paucity of more rapidly spinning systems has serious
implications for the prospects for future detections and corresponding
constraints on the neutron star EOS.

To explore these implications further, it is worthwhile to consider how
much faster a neutron star need be discovered before significant
constraints on the possible EOS are achieved. In this respect the
candidate 1122~Hz burst oscillation,
even though it has not been
confirmed, has prompted a timely exploration of the consequences for the EOS.
According to \cite[]{lavagetto06}, this result leads for the
first time to ``strong, model-independent observational constraints'' to the
neutron star EOS (see also \citep{bhz07}).  For
such a rapidly spinning neutron star to be comprised of nucleonic matter
would require a rather large mass of $\gtrsim2~M_\odot$ (perhaps providing
an alternative explanation of why extremely rapidly-spinning neutron stars
have been so hard to find). The rotational mass-radius limit for an
1122~Hz neutron star just intersects the M-R trajectories for several
plausible equations of state at the highest possible mass (\citep{lp07},
Figure 2). This suggests
that this spin rate is a convenient empirical target for observers;
neutron stars spinning slower than this rate likely cannot
significantly constrain the EOS (unless other, complementary constraints
are available) while neutron stars spinning at even higher rates have a good
chance to constrain the EOS.

It is also worth noting that the prospect for access to an X-ray
timing mission in the near future is far from guaranteed. The present
\xte\/ cycle 12, through February 2009 at the latest, may be the last
observing cycle for the instrument\footnote{URL \url{
http://heasarc.gsfc.nasa.gov/docs/xte/cycle12.html}}. Efforts are underway
to continue the mission through 2009 and beyond, but if these efforts are
unsuccessful, no alternative timing mission is currently planned for the
near future by ESA or NASA. The best chance for a replacement
high-sensitivity, high time resolution X-ray instrument is the Large-Area
Xenon Proportional Counter (LAXPC) onboard the Indian multiwavelength
satellite {\it ASTROSAT}\footnote{URL
\url{http://meghnad.iucaa.ernet.in/~astrosat}}, currently scheduled for
launch in December 2009. The LAXPC has comparable spectral and timing
resolution to the \xte/PCA, with improved high-energy sensitivity; in
addition, the satellite will also feature soft- and hard-X-ray imaging
telescopes, an all-sky monitor, and a UV telescope for broadband coverage.

In the unfortunate event that \xte\/ ceases operation before {\it
ASTROSAT}\/ is launched, there will be no further spin measurements 
for rapidly-rotating accreting neutron stars in the meantime. Even if this
situation is avoided, if some phenomenon prevents the spin-up of neutron
stars to spin rates much in excess of 750~Hz (as is suggested by the
present distribution of measured spins), the prospects for strong
constraints on the neutron star EOS by measurement of rapid spins alone
appear poor. However, the prospects for constraints via multiple
observational measurements remain promising. For the accretion-powered
neutron stars, this approach is illustrated by the recent results
from EXO~0748$-$676, which combined the spin rate with measurements of the
surface gravitational redshift, the peak flux of radius-expansion
thermonuclear bursts (the Eddington limit) and the (apparent) blackbody
radius of the star from the X-ray flux in the burst tail \cite[]{ozel06}.
Although the spin rate in this system is the slowest measured in any LMXB
at 45~Hz, and so cannot alone give any useful constraints on the EOS, the
combination of other measurements allowed those authors to rule out all
the ``soft'' equations of state for this system.

Although this result is not without it's own caveats (see e.g.
\citep{alford06}),
many of the issues appear resolvable.
The energetics of both thermonuclear bursts and carbon-burning
``superbursts'' may also allow complementary measurements of the heat flux
from the neutron star crust, which also constrains the interior properties
and hence the EOS (e.g. \citep{pc05}).

\begin{table}
\begin{tabular}{lcccccl}
\hline
       &      & $\nu_{\rm spin}$ & $P_{\rm orb}$ & $d$ & 
$\dot{M}$ & \\
Source & Type\tablenote{B = burster, T = transient, L = long-duration ($>1$~yr) 
transient, D = dipping source. Adapted from \cite[]{lmxb07}.} &  (Hz) &
(min)        & (kpc)\tablenote{For the sources with radius-expansion
bursts, the distance is determined from the mean peak flux of those bursts
\cite[]{bcatalog}. For sources with only
non-radius expansion bursts, the maximum peak flux of the bursts leads to
an upper limit on the distance. For sources with no bursts, the ratio of
the long-term average X-ray flux and the expected accretion rate driven by
gravitational radiation leads to a lower limit on the distance
\cite[]{gal06b} }
& ($10^{-11}\
M_\odot\,{\rm yr^{-1}}$)\tablenote{The inferred mass accretion rate, based on the
time-averaged broadband X-ray flux and the distance estimate, and assuming
a $1.4M_\odot$ neutron star with radius 10~km. For
transients, the peak rate observed during outburst is indicated in
parentheses. For the long-duration transients, the rate is
measured while the source is X-ray active. For sources without bursts, the
estimated rate is a lower limit based on the minimum companion mass and
the orbital period (following \citep{bc01}).}
 & Ref.\\
\hline
\multicolumn{7}{c}{Burst oscillation sources}\\
\hline
EXO 0748$-$676	& BDT	& 45	& 229 & 7.5 & 20--45 (120) & \cite{villarreal04} \\
4U 1916$-$05	& BD	& 270	& 50 & 8.9 & 22--110 & \cite{1916burst} \\
IGR J17191$-$2821	& BT	& 294	& $\ldots$ & $<11$ & $\ldots$ & \cite{mark07a} \\
4U 1702$-$429	& B	& 329	& 1320 & 5.5 & 35--80 & \cite{mss99} \\
4U 1728$-$34	& B	& 363	& $\ldots$ & 5.2 & 45--210 & \cite{stroh96} \\
KS 1731$-$26	& BL	& 524	& $\ldots$ & 7.2 & 85--350 & \cite{muno00} \\
A 1744$-$361	& BT	& 530	& $\ldots$ & $<9$ & ($<230$) & \cite{bhatt06b} \\
MXB 1658$-$298	& BDL	& 567	& 427	& 12 & 60--200 & \cite{wij01} \\
4U 1636$-$536	& B	& 581	& 228	& 6.0 & 55--330 & \cite{zhang97} \\
GRS 1741.9$-$2853\tablenote{The source of the bursts with 589~Hz
oscillations assumed by \cite{bcatalog}. } & BT	& 589	& $\ldots$ & 8 &
0.01 (1) & \cite{stroh97,muno03b} \\
SAX J1750.8$-$2900 & BT	& 601	& $\ldots$ & 6.8 & 20 (180) & \cite{kaaret02} \\
4U 1608$-$52\tablenote{Hartman et al. 2008, in preparation}	& BT	&
  620	& 773	& 4.1 & 100 (530) & \\
\hline
\multicolumn{7}{c}{Accretion-powered millisecond pulsars} \\
\hline
Swift J1756.9$-$2508 & T & 182 & 54.0	& 8? & $>0.14$ (200) & \cite{krimm07} \\
XTE J0929$-$314 & T & 185 & 43.6 & $>4$	& $>0.4$ (30) & \cite{gal02d} \\
XTE J1807$-$294 & T & 191& 40.1 & $>5$ & $>0.3$ (40) & \cite{markwardt03b} \\
XTE J1814$-$338\tablenote{Also a burst oscillation source} & BT & 314 &
  257 & $<8$ & 1.4--1.6 ($<60$) & \cite{stroh03a} \\
SAX J1808.4$-$3658$^\mathparagraph$ & BT	& 401 & 120.9 & 3.4--3.6 &
  0.6--1.3 (75) & \cite{wij98b,chak98d,gal06c} \\
XTE J1751-305 & T & 435	& 42.4	& $>8$ & $>1.2$ (190) & \cite{markwardt02} \\
IGR 00291+5934 & T & 599 & 147.4 & 5--6 & 1.3 (90) & \cite{gal05a,gal06b} \\
\hline
\multicolumn{7}{c}{Intermittent pulsars}\\
\hline
HETE J1900.1-2455 & BL & 377 & 83.3 & 5.0 & 7--70 & \cite{kaaret05b,gal07a} \\
SAX J1748.9$-$2021 & BT & 442 & 520 & 8.1 & 8 (420) & \cite{gavriil07,altamirano07} \\
Aql X-1$^\mathparagraph$ & BT & 550 & 1137 & 5.0 & 105 (510) & \cite{casella07} \\ 
\hline
\end{tabular}
\caption{Rapidly-rotating accreting neutron stars}
\label{tab:a}
\end{table}

\begin{theacknowledgments}
Rudy Wijnands provided timely and insightful comments on
this paper. I would also like to thank the conference
organizers for a highly entertaining and enlightening meeting.
\end{theacknowledgments}

\end{document}